\documentstyle[prd,aps,preprint]{revtex}
\tighten

\def\S{{\cal S}}
\def\be{\begin{equation}}
\def\ee{\end{equation}}
\def\bea{\begin{eqnarray}}
\def\eea{\end{eqnarray}}

\renewcommand\({\left(}
\renewcommand\){\right)}
\renewcommand\[{\left[}
\renewcommand\]{\right]}

\newcommand\eq[1]{Eq.~(\ref{#1})}

\newcommand\eqsss[4]{Eqs.~(\ref{#1}), (\ref{#2}), (\ref{#3})
and (\ref{#4})}

\newcommand\TeV{\,\mbox{TeV}}
\newcommand\GeV{\,\mbox{GeV}}
\newcommand\MeV{\,\mbox{MeV}}
\newcommand\keV{\,\mbox{keV}}

\newcommand\msun{M_\odot}

\newcommand\lsim{\mathrel{\rlap{\lower4pt\hbox{\hskip1pt$\sim$}}
    \raise1pt\hbox{$<$}}}
\newcommand\gsim{\mathrel{\rlap{\lower4pt\hbox{\hskip1pt$\sim$}}
    \raise1pt\hbox{$>$}}}

\newcommand\diff{\mbox d}

\def\calp{{\cal P}}

\def\calv{{\cal V}}

\newcommand\bfx{{\bf x}}

\newcommand\sub[1]{_{\rm #1}}

\newcommand\cdm{\sub{cdm}}

\begin{document}
\preprint{astro-ph/0306500}

%
%
\input epsf
%

\title{The CDM isocurvature perturbation in the curvaton scenario}
\author{David H.\ Lyth$^1$ and David Wands$^2$}
\address{(1) Department of Physics, Lancaster University, Lancaster
LA1 4YB,~~~U.~K.}
\address{(2) Institute of Cosmology and Gravitation, University of
Portsmouth, Portsmouth PO1 2EG,~~~U.~K.}
\date{\today}
\maketitle
\begin{abstract}
  We discuss the residual isocurvature perturbations,
  fully-correlated with the curvature perturbation, that are
  automatic in the curvaton scenario if
  curvaton decay is  sufficiently late. We contrast these residual
  isocurvature perturbations with the
  generally un-correlated `intrinsic' isocurvature perturbation
  generated by an additional field such as the axion.  We present a
  general formula for the residual isocurvature perturbations,
  referring only to the generation of the relevant quantity (Cold Dark
  Matter, baryon number or lepton number) in an {\em unperturbed}
  universe. Specific formulas for the residual isocurvature CDM
  perturbation are given, for most of the commonly-considered CDM
  candidates.
\end{abstract}

\pacs{PACS numbers: 98.80.Cq\hfill Preprint astro-ph/0306500, PU-ICG-03/11}



\section{Introduction}

{}The inhomogeneity of the early Universe first becomes measurable
a few Hubble times before cosmological scales enter the horizon.
At this stage, well after Big Bang Nucleosynthesis (BBN), we know
that the Universe consists primarily of photons, neutrinos,
baryonic matter and Cold Dark Matter (CDM) \cite{KT}. The perturbations in
the densities of these components are conventionally specified by
the perturbation in the total energy density, characterised by a
spatial curvature perturbation, $\zeta$, and by three isocurvature
perturbations
\be
 \S_i \equiv \frac{\delta (n_i/n_\gamma)}{ (n_i/n_\gamma) }
 \label{sidef1} \,,
\ee
with $i=$CDM, $B$ or $\nu$,   giving the inhomogeneity in the
relevant number density per photon \cite{book}.
As we shall see,  $\zeta$ and
the $\S_i$ are constant from BBN to the approach of horizon entry.
They set the initial conditions for the subsequent evolution of
cosmological perturbations, and we refer to them collectively as
the `Primordial Density Perturbation'.

Observation is {\em consistent} with the hypothesis that the
isocurvature perturbations are absent, while the curvature
perturbation is Gaussian with an almost perfectly flat spectrum
\cite{wmapspergel,wmappeiris,wmapng}.
On the other hand, spectral tilt, non-Gaussianity and isocurvature
perturbations may all be present at some level.

The Primordial Density Perturbation presumably originates during
inflation, from the vacuum fluctuation of one or more light scalar
fields. According to the usual hypothesis \cite{book}, it originates
solely from the vacuum fluctuation of the inflaton field, defined as
the one whose value determines the end of inflation. This `inflaton
scenario' leads to definite expectations about the nature of the
primordial density perturbation. There are no isocurvature
perturbations, and non-gaussianity will almost certainly exist
\cite{infng} only at the $10^{-5}$ level through second-order
perturbations.\footnote {Bigger non-gaussianity of the $\chi^2$ type
  can be generated in a multi-field model, but only at the expense of
fine-tuning \cite{multiinfng}.}  On the other hand,
significant spectral tilt is expected, and significant running of the
spectral index is quite possible \cite{LRreview,book}.
It is also possible
that the curvature perturbation is accompanied by gravitational
waves of sufficient amplitude to affect the Cosmic Microwave
Background (CMB) anisotropy \cite{LRreview,book}.
These expectations have achieved the
status of a `standard model' of the early Universe, which is now
used by almost all groups who analyse CMB and galaxy distribution
measurements.

It has been pointed out recently \cite{lw} (see also \cite{sylvia,lm})
that the inflaton scenario
is not the only possible mechanism to generate large scale structure from
inflation.
The primordial density perturbation may instead originate from the
vacuum fluctuation of some `curvaton' field, different from the
inflaton field. The curvaton energy density has an isocurvature
perturbation, and this generates the curvature perturbation when the
curvaton density comes to be a significant fraction of the total
before the curvaton finally decays.\footnote
{Recently, a completely different
scheme has
been proposed \cite{dgz,kofman,emp},  in which  the responsible field
acts by perturbing the inflaton decay rate  without ever contributing
significantly to the energy density.}
This `curvaton scenario'
leads to completely different expectations about the nature
of the primordial density perturbation.  Gravitational waves
\cite{dl} (see also \cite{wbmr}), and probably also the spectral tilt
 will be negligible, but 
non-Gaussianity and isocurvature
perturbations are perfectly possible \cite{luw}.

The curvaton scenario has received  
a lot of attention \cite{dl,luw,mt1,andrew,mt2,fy,hmy,hofmann,bck1,bck2,%
mormur,ekm,mwu,postma,fl,gl,kostasquint,giov,lu,mcdonald,dllr1,ejkm,dlnr,%
ekm2,pm,kkt,lw03,dllr2}
because it opens up new possibilities both for
model-building and for  observation.
The present paper focusses on the possible isocurvature
perturbations, following \cite{luw}. Because they have the same
source as the curvature perturbation, they  are fully correlated
 \be
\S_i(\bfx) = s_i \zeta(\bfx)
\,,
\label{residual}
 \ee
with $s_i$ numbers that depend on the physical model. Following
\cite{luw}, we  call this kind of isocurvature perturbation {\em
residual}. A given $s_i$ will be zero if the relevant quantity
(CDM, baryon number or lepton number) is created after the
curvaton decays. Otherwise it can be calculated within a given
early-Universe scenario, and it is generally expected to be of
order unity. Our main purpose is to derive specific formulas for
the CDM case, for the various possibilities regarding the nature
of the CDM that are commonly envisaged.

In either the inflaton or the curvaton scenario, one can suppose
that the vacuum fluctuation of one or more additional fields is
also relevant. Such additional fields may generate additional
isocurvature perturbations, which will in general be {\em
uncorrelated} with the curvature perturbation and at best only
partially correlated. We shall call these {\em intrinsic}
isocurvature perturbations. In general, intrinsic isocurvature
perturbations and the curvature perturbation are determined by
different sectors of the underlying particle theory. This means
that in general, there is no reason to expect the magnitude of an
intrinsic isocurvature perturbation to be comparable with that of
the curvature perturbation. In a particular model though, it may
turn out that a single sector of the theory accounts for both the
curvature perturbation and an intrinsic isocurvature perturbation,
so that their magnitudes are comparable. (See \cite{dllr1} for the
only example known so far, arising within the curvaton scenario.)

The plan of the paper is as follows. In Section \ref{conserved} we
recall the general description of the Primordial Density Perturbation, 
in terms of quantities
which are conserved on super-horizon scales. In Section
\ref{intres} we  discuss how isocurvature perturbations can
originate, first  in the inflaton scenario and then in the
curvaton scenario, giving a general formula for the residual
isocurvature perturbations. In Section \ref{residualsec} we apply
our formula,  to evaluate the residual CDM isocurvature
perturbation for each of the commonly-envisaged CDM possibilities.
We conclude in Section \ref{conclusion}.

\section{Conserved perturbations}

\label{conserved}

The discussion of  the evolution of perturbations
while they are outside the horizon
is facilitated by the existence of perturbations that are,
under suitable circumstances, conserved \cite{wmll,lw03}.
The conserved perturbations are conveniently defined with
reference to the slices of spacetime which have zero intrinsic
curvature perturbation (spatially flat slices), because on
super-horizon scales the local expansion of the Universe between
such slices is uniform \cite{wmll,lw03}. By virtue of this
uniformity, a perturbation will be conserved on super-horizon
scales if it is constructed according to the following recipe
\cite{lw03}. Consider a quantity $f(\bfx,t)$, with $t$ the
coordinate time labelling spatially flat slices. If $f$
decreases (or increases) monotonically with
the comoving volume $\cal V$ according to an equation of the form
\be
{\cal V}\frac{\partial f}{\partial \cal V} = y(f)
 \label{calv}
\,,
\ee
then the perturbation
\be
 X_f \equiv -H\frac{\delta f}{\dot f} \label{defXg}
 \ee
is conserved \cite{lw03}, where $\delta f$ is to be evaluated on
spatially flat slices. In practice we find conserved quantities of
three types.

\paragraph{The total energy density perturbation.}
The quantity
\be
\zeta \equiv  -H \frac {\delta\rho}{\dot\rho}
=  \frac {\delta\rho}{\rho+ P}
\label{zetadef}
\ee
is conserved provided that the pressure $P$ of the Universe is a unique
function of the energy density $\rho$, or equivalently if the perturbations
satisfy the adiabatic condition
\be
\delta P/\delta \rho = \dot P/\dot \rho
\label{padiabatic}
\,.
\ee
More general, if the pressure is non-adiabatic,
\be
\dot\zeta =   -\frac H{\rho + P} \delta P\sub{nad}
\,,
\ee
where
\be
\delta P\sub{nad} \equiv \delta P - \frac{\dot P}{\dot \rho}
\delta \rho
\label{pnaddef}
\,.
\ee

\paragraph{Separate energy density perturbations.}
Let  $\rho_i$ and $P_i$ refer to some component which by itself
satisfies the adiabatic condition \eq{padiabatic}, and which does not exchange
energy with any other component. Then the quantity
\be
\zeta_i \equiv  -H \frac {\delta\rho_i}{\dot\rho_i} \label{zetaidef}
=  \frac {\delta\rho_i}{\rho_i+ P_i}
\ee
is conserved. If the component is radiation ($P_i=\rho_i/3$),
\be
\zeta_i \equiv   \frac14 \frac{\delta\rho_i}{\rho_i}\ \ \ {\rm (radiation)}
 \label{zetarad}
\,,
\ee
and if it is matter ($P_i=0$)
%
\be
 \zeta_i \equiv  \frac13  \frac{\delta\rho_i}{\rho_i}\ \ \ {\rm
(matter)} \,.
\ee
The conservation of $\zeta_i$ in these cases is an immediate
consequence of the dependence of the local energy densities on
comoving volume $\calv$ (namely
$\rho_i\propto \calv^{-1}$ for
matter and $\zeta_i\propto \calv^{-4/3}$ for radiation).

\paragraph{Number density perturbation.}
Let  $n_i$ be any conserved number density. Then, since $n_i\propto
\calv^{-1}$ the quantity
 \be
\tilde\zeta_i \equiv \frac13 \frac{\delta n_i}{n_i}
 \label{tzi} \,,
 \ee
is conserved.

\subsection{The primordial density perturbation}

The  epoch $T\sim 10\MeV$, marking the beginning of of the
Big-Bang Nucleosynthesis process, is the earliest one of which we
have definite knowledge. After $T$ falls below $1\MeV$, the energy
density
is dominated by radiation in the form of
photons and neutrinos, and there is also baryonic matter and
presumably Cold Dark Matter (CDM);\footnote {If  non-relativistic
particles decaying well after nucleosynthesis have a significant
effect, we are assuming here that the perturbation in their energy
density is adiabatic and hence need not be considered separately.}
The number of particles is conserved for each component, and there
is no exchange of energy between components. As a result the
following quantities are conserved until the approach of horizon
entry:
\bea
\zeta\cdm &\equiv &   \frac13\frac {\delta\rho\cdm}{ \rho\cdm} \label{zcdm}\\
\zeta_B &\equiv &  \frac13\frac {\delta\rho_B}{ \rho_B} \label{zb}\\
\zeta_\nu &\equiv &   \frac14\frac {\delta\rho_\nu}{ \rho_\nu} \label{znu} \\
\zeta_\gamma &\equiv &   \frac14\frac {\delta\rho_\gamma}{ \rho_\gamma}
\label{zgamma}
\eea

The  three isocurvature perturbations \eq{sidef1} are given by
\be
 \label{defSi}
\S_i \equiv 3(\zeta_i - \zeta_\gamma)
\,,
\ee
and they too are conserved until the approach of horizon entry.
The total energy density/curvature perturbation, \eq{zetadef}, is
given by
 \bea
\zeta &=&-H \frac{\sum \rho_i \zeta_i}{\dot \rho} \label{zetasum}\\
& \simeq& (1-f_\nu) \zeta_\gamma + f_\nu \zeta_\nu \label{zetaprimord}
\,,
 \eea
during the primordial era when the matter density is negligible
and $f_\nu =\rho_\nu/\rho_\gamma$ is a constant. The curvature
perturbation $\zeta$ is thus constant during the radiation
dominated era to high accuracy until the approach of horizon
entry.

The smallest cosmological scale, enclosing say  $M\sim 10^6\msun$,
enters the horizon at $T\sim 1\keV$. The curvature perturbation
$\zeta$ and the isocurvature perturbations $\S_i$ therefore have
constant values on all cosmological scales in the regime $1\keV
\lsim T \lsim 1\MeV$. We are calling them collectively the
Primordial Density Perturbation, and through the Einstein field
equation and the Boltzmann hierarchy, they specify the subsequent
linear evolution of the entire set of cosmological perturbations
\cite{book,bmt}. They are therefore determined by observation.

It is found from observation \cite{wmapspergel,wmappeiris,wmapng} that
the primordial curvature perturbation is almost Gaussian, with an
almost scale-independent spectrum given by $\calp_\zeta^\frac12 \simeq
5\times 10^{-5}$. Current observations provide only upper bounds on
the isocurvature perturbations
\cite{trotta,amendola,gl,wmappeiris,finns,juan}, which depend on the
degree of correlation between them and the curvature perturbation. In
this paper we are concerned with the fully-correlated perturbations
that can be produced only in the curvaton scenario. For them, the
ratios $\S_i(\bfx)/\zeta(\bfx)$ are constants determined by the
particle physics model. 

The observational bound in the case of $s\sub
B=\S\sub B(\bfx)/\zeta(\bfx)$ is \cite{gl} at $95\%$ confidence level
 \be
 -0.53  < s\sub B < 0.43 \,.
 \ee
The bound on $s\cdm=S\cdm(\bfx)/\zeta(\bfx)$ is tighter by
precisely the factor $\Omega\sub B/\Omega\sub{cdm}$ \cite{gl}, and
taking that ratio as $1/6$ we get
 \be
  -0.09 < s\cdm < 0.07 \,.
\label{scdmbound} \label{observcdm}
 \ee
The observational bound on $s_\nu$ is \cite{juan}
\be
-0.14 < s_\nu < 0.47 
 \label{snubound}
\,.
\ee
This bound has been calculated for negligible lepton asymmetry, whereas 
according to \eq{snu} below 
a non-zero lepton asymmetry is required \cite{luw}
 to produce an neutrino isocurvature density
perturbation.
Nevertheless, because the lepton asymmetry is small the bound
may be expected to provide a good approximation.

\subsection{Conserved number densities}

To describe the origin of the isocurvature perturbations in the
early Universe, we need to consider \cite{luw,lw03} the CDM number
density $n\cdm$, the density of baryon number $n_B$ and the
density of lepton number $n_L$. Their perturbations may be
characterised by \cite{luw}
 \be
\tilde\zeta_i \equiv   - H\frac{\delta n_i}{\dot n_i}
 \label{tzcdm}
\,,
 \ee
with $i=$cdm, $B$ or $L$. We can take the epoch of creation of
CDM, baryon number or lepton number to be the epoch after which
the relevant quantity (CDM particle number, baryon number or
lepton number) is conserved. Each quantity $\tilde \zeta_i=\delta
n_i/3n_i$ is then conserved after the relevant epoch of creation
\cite{lw03}.

In the case of CDM, $\zeta\cdm$ in \eq{zcdm} and $\tilde
\zeta\cdm$ \eq{tzcdm} will usually be equal from the moment of CDM
creation, and in any case are equal by the primordial epoch. The
exception is the case of axionic CDM though, where the axion mass
is initially temperature-dependent so that only $\tilde\zeta\cdm$
is conserved after creation. In that case, $\zeta\cdm$, will be equal to
$\tilde\zeta\cdm$ after the mass becomes constant.

It follows from Eqs.~(\ref{defSi}) and~(\ref{zetaprimord}) that
 \bea
\S\cdm &=& 3(\tilde\zeta\cdm-\zeta_\gamma) \\
&=& 3(\tilde\zeta\cdm-\zeta)
\,.
 \eea
The final equality assumes that $\S_\nu=0$, but the
 modification
in the  case of nonzero $\S_\nu$ is quite straightforward \cite{luw}
and there is no need to consider it here.

In the case of baryonic matter, there is no useful definition of
$\zeta_B$ before the quark-hadron transition, since baryon number
is not carried by a particular energy density. Afterwards,
$\zeta_B=\tilde\zeta_B$ leading to
 \bea
\S\sub B &=& 3(\tilde\zeta\sub B-\zeta_\gamma) \\
&=& 3(\tilde\zeta\sub B-\zeta)
\,.
\eea

In the case of neutrinos, $\zeta_\nu$ is of course always defined
provided that neutrinos exist but it is a useful quantity only
after neutrinos freeze out of thermal equilibrium. After that
epoch \cite{luw},
 \bea
 \S_\nu &=& \frac{135}{7} \( \frac\xi\pi \)^2 \( \tilde \zeta_L
-\zeta_\gamma \) \\
&\simeq &  \frac{135}{7} \( \frac\xi\pi \)^2 \( \tilde \zeta_L
-\zeta\)
\,.
\label{snu}
 \eea
In these expressions,
 $\tilde\zeta_L$ is the perturbed net lepton number, $\delta
n_L/3n_L$, and $\xi$ is the neutrino asymmetry parameter,
constrained by nucleosynthesis to $|\xi|<0.07$. The second equality
is a good approximation \cite{luw} by virtue of the bound on $|\xi|$.

In the curvaton scenario, $\tilde\zeta_L$ is negligible compared with
$\zeta$ if the neutrino
asymmetry is created before the curvaton contributes significantly 
to the energy density \cite{luw}. Then, the present observational
bound \eq{snubound} leads to $-\xi < 0.27$, which is competitive
with the nucleosynthesis bound. 

\section{Intrinsic and residual isocurvature perturbations}

\label{intres}

\subsection{The inflationary initial condition}


Inflation is supposed to set the initial condition for the
subsequent evolution of the Universe, through the values of the
light scalar fields which exist at the end of inflation. We are
here defining a field as light if the second derivative of the
potential in that direction is much less than $H^2$. This means
that a light field has a flat spectrum of perturbations with
spectrum $(H_*/2\pi)^2$, with the star denoting the epoch of
horizon exit.

To handle the perturbed Universe, we adopt what has been called
the separate universe picture \cite{wmll,book}, which means that
the local evolution of regions within our presently observable
Universe is the same as that of an unperturbed FRW
universe.
One assumes
that the Universe is smoothed on the shortest relevant scale. At
least for cosmological scales the separate universe picture seems
very likely to be valid, since such scales are far larger than the
Hubble scale in the very early Universe.\footnote
{In \cite{wmll,book} the separate universe picture at a given epoch
is taken to be valid for any smoothing scale bigger than the Hubble
distance. That is a stronger assumption, which might not be valid 
in the presence of short-distance phenomena like preheating 
\cite{llmw}.}
 As a result, physical
processes in the very early Universe are unlikely to distinguish
between the shortest cosmological scale (say, $1$~Mpc) and the
longest one corresponding to the present Hubble scale
($\sim10^4$~Mpc). And the separate universe assumption is
certainly valid on the latter scale, or else the whole concept of
an using an unperturbed FRW background to describe our observable
Universe makes no sense.

For an unperturbed FRW universe, a generic quantity $g$ (such as
energy density, pressure, number density or a scalar field) has an
evolution of the form
\be
 g(\phi_1,\phi_2,\cdots,N) \,
\label{unpert}
\,,
 \ee
where $\phi_i$ are the values of the light fields specified at
some initial epoch.
We are specifying the epoch at which $g$ is to be evaluated by the
amount of expansion
\be
N \equiv \frac13\ln({\cal V}/{\cal V}\sub{initial})
\equiv \ln(a/a\sub{initial}) \equiv \int^t_{t\sub{initial}}  H \diff t 
\,.
\label{nexp}
\ee
In the separate universe picture, the locally-defined generic
quantity $g$ has the same functional form as in the unperturbed
Universe, \eq{unpert}, but with position-dependent arguments;
\be
 g(\phi_1(\bfx),\phi_2(\bfx),\cdots,N(\bfx,t)
) \,,
 \label{pert}
\ee
where $(\bfx,t)$ are the coordinates (gauge) used to describe the
perturbations, and $N$ is the locally-defined defined quantity.\footnote
{The final expression of  \eq{nexp} is now $\int H\diff \tau$
where $H$ is the locally-defined quantity Hubble expansion and
$\tau(t)$ is the local proper time.}

The first-order perturbation can be written as
\be
 \delta g
 = \sum_i \frac{\partial g}{\partial \phi_i} \delta\phi_i(\bfx)
  + \frac{\partial g}{\partial N} \delta N(\bfx,t)
 \,.
\label{deltag}
 \ee
 For a given wavenumber $k/a$ one can choose the initial time to be a
 few Hubble times after horizon exit during inflation. Then the
 $\delta\phi_i(k)$ are uncorrelated gaussian quantities, with an
 almost flat spectrum
\be
 \calp(k) = (H_*/2\pi)^2 \,,
 \label{specphi}
\ee
where the star denotes the epoch of horizon exit $aH=k$.

Any conserved quantity $X_f$, defined in \eq{defXg}, can be
calculated by evaluating $\delta f$ on a spatially flat time-slice
as a function of the initial field perturbations . This is
separated from an initial spatially flat time-slice during
inflation by a uniform expansion, i.e., $\delta N=0$, which yields
from \eq{deltag}
\be
 X_f = -\frac{H}{\dot{f}} \delta f
  = -\frac{H}{\dot{f}} \( \frac{\partial f}{\partial \phi_i} \)_N
   \delta \phi_i(\bfx) \,,
\ee
In particular we can write the total curvature perturbation as
%
\be
 \zeta = -\frac{H}{\dot\rho} \delta \rho
  = -\frac{H}{\dot\rho}
   \sum_i \( \frac{\partial \rho}{\partial \phi_i} \)_N\delta \phi_i(\bfx) \,,
 \label{calczeta}
\ee

This is a generalisation of the approach advocated by Sasaki and
Stewart \cite{ewanmisao} as the simplest way to calculate the
curvature perturbation in multi-field models.  They calculated
$\zeta=\delta N$ in the particular case where $N$ is the local
expansion between an initial spatially flat time-slice and a final
uniform-density slice \cite{wmll}. Taking $g=\rho$ and requiring
$\delta\rho=0$ in \eq{deltag} reproduces this expression:
%
\be
 \zeta = \delta N = -\sum_i \left( \frac{\partial N}{\partial
  \phi_i} \right)_\rho  \delta\phi_i(\bfx)
 \,.
\label{deltaN}
 \ee
where we have used $dN/d\rho=H/\dot\rho$ along the background
solution. 

The general expression \eq{deltag} has not been written down
before.
But it is all that is
required in the separate universe picture to calculate the
conserved quantities $\zeta_i$ \eq{tzcdm} produced from field
fluctuations during inflation, and hence describe the Primordial
Density Perturbation.
In what follows we will be interested in the case of isocurvature
perturbations corresponding to conserved number densities $n_i$,
and our main result will be \eq{zetaigeneral}, which along with
\eq{saxion} is a further example of \eq{deltag}.

\subsection{The inflaton scenario}

During single-field slow-roll inflation,\footnote {We adopt the
now-standard terminology, where `single-field' means that there is
an essentially  unique slow-roll trajectory while `multi-field'
means that there is a  family of slow-roll trajectories.} the
inflaton field satisfies $3H\dot\phi=-V'(\phi)$, which can be
integrated to give $\phi$ as a unique function $\phi(N)$ of
the integrated expansion $N$, up to a constant of integration.

In the the inflaton scenario, $\phi$ sets the initial condition for
both the energy density and the pressure.
As a result, the pressure
perturbation is adiabatic, making $\zeta$ constant. Using
\eq{zetadef} with $\delta \rho \simeq V'\delta\phi$  one finds that a
few Hubble-times after horizon exit the well-known
formula
 \be
\zeta \simeq  -\frac{H}{\dot\phi} \delta \phi \label{zetainf} \,,
 \ee
where the right hand side can be evaluated at horizon exit.
In the inflaton scenario, this value is maintained until the primordial
epoch.

A primordial isocurvature perturbation $\S_i$ can
be generated in the inflaton
scenario if the relevant number density $n_i$ at the time of its
creation depends on some scalar field $\chi$ as well as the inflaton
field;
\be
 n_i(\phi(\bfx),\chi(\bfx),N(\bfx)) \,.
\ee
%
We shall call this kind of isocurvature perturbation an {\em
intrinsic} one, as opposed to the `residual' one that can appear
only in the curvaton scenario. The known cases where it is
possible are when  the CDM comes initially from an oscillating
field (axion or Wimpzilla), or when there is Affleck-Dine baryo-
or lepto-genesis. The qualification `possible' is here crucial,
because the axion or Affleck-Dine field has to be light during
inflation so as to acquire a perturbation, and also because that
perturbation has to survive until $n_i$ is created. This does not
happen if the CDM axions come mostly from the oscillation of
Peccei-Quinn cosmic strings, or if the Affleck-Dine field has the
generic effective mass-squared  of order $\pm H^2$.

The possibilities for obtaining an intrinsic isocurvature
perturbation are  very limited, so that, for example, there is no
known way of obtaining an intrinsic CDM isocurvature perturbation
if the CDM consists of the relic abundance of the Lightest
Supersymmetric Particle obtained when it decouples from thermal
equilibrium. Even when there is an intrinsic isocurvature
perturbation, there is in general no reason to expect its
magnitude is comparable with that of the adiabatic perturbation.
This is because the magnitudes  of the two perturbations depend on
different and in general unrelated parameters, as a result of the
fact that the perturbations are caused by different fields.

A final and very important property of an intrinsic isocurvature
perturbation is that it is uncorrelated with the adiabatic
perturbation. Again, this comes from the fact that two different
fields are involved.

With one possible exception, the above discussion applies also to
multi-field inflation, if we define the inflaton as the specific
linear combination of fields corresponding to the inflaton
trajectory at the {\em end} of inflation. The possible exception
concerns the correlation; there is correlation in the exceptional
case that the combination of fields orthogonal to the inflaton was
responsible for the isocurvature perturbation \cite{langlois}.
However, no example of this has been proposed in the context of
particle physics; in particular, no model has been exhibited where
this combination is the axion or an Affleck-Dine field.

\subsection{The  curvaton scenario}

In the   curvaton scenario,  the value \eq{zetainf} of $\zeta$
generated by the inflaton is supposed to be negligible compared
with the observed value of order $10^{-5}$, and to remain
negligible until the curvaton field $\sigma$ starts to oscillate.
The
density $\rho_\sigma$ is supposed to be negligible when the
oscillation starts, but during radiation domination it grows like
$a(t)$ relative to the radiation density
Until the curvaton starts to decay, the energy densities of the
radiation and the curvaton are
decoupled,
 so that 
$\zeta\sub r$ is constant
and negligible, while $\zeta_\sigma$ is constant but not
negligible. {}From \eq{calczeta} we then have
 \be
 \zeta(t,{\bfx}) = -\frac{H}{\dot\rho}
  \left(\frac{\partial\rho}{\partial\sigma}\right)
  \delta\sigma({\bfx})
= f(t) \zeta_\sigma({\bfx}) \label{zetaoff}
 \label{zetat}
\,,
 \ee
where $\zeta_\sigma=-H\delta\sigma/\dot\sigma$ and
 \be
f \equiv \frac{\dot\rho_\sigma}{\dot\rho} =
 \frac{3\rho_\sigma}{4\rho\sub r + 3 \rho_\sigma} \label{fdef} \,,
 \ee
is a growing function of time.

When curvaton decay is complete, the Universe is supposed to be thermalised,
so that the temperature determines the entire future evolution 
of the energy density and pressure. (Non-thermalised components such as
CDM are supposed to give a negligible contribution to these quantities
until at least the primordial epoch.)
As a result, the pressure in each separate universe is
once again a practically unique function of energy density,
leading to a practically constant value of
$\zeta$ given by \eq{suddec}.
Following \cite{luw,mwu} we define
 \be
 \label{defr}
r\equiv \zeta / \zeta_{\sigma} \,,
 \ee
where $\zeta$ on the left-hand-side is evaluated after curvaton
decay is complete, and $\zeta_\sigma$ on the right-hand-side is
evaluated before the decay starts.

In the sudden decay approximation we have
 \be
 \zeta \simeq f(t\sub{dec}) \zeta_\sigma
 \label{suddec}
\,.
 \ee
Hence we have $r\sim(\rho_\sigma/\rho)\sub{dec}$.
Equation~(\ref{suddec}) gives a good approximation (to within
10\%) to the precise numerical solution \cite{mwu}. Note that $r$
is found to be always less than (or equal to) one and, in
particular, $r\to 1$ in the limit $\rho_\sigma/\rho\to 1$.

In the curvaton scenario there might be an intrinsic
primordial isocurvature
perturbation due to some other field $\chi$, different from both
the inflaton $\phi$ and the curvaton $\sigma$. If present, it will
have the features discussed already in the case of isocurvature
perturbations for the inflaton scenario. It will generally be
uncorrelated with the residual isocurvature perturbation and
accordingly we ignore it from now on.

We are interested in a different type of primordial isocurvature
perturbation, which we term residual \cite{luw}. A residual
isocurvature perturbation  can arise because in the curvaton
case, the separate universes are not identical while the curvaton
is oscillating. This is because they are characterised by the two
independent local energy densities $\rho_\phi$ and $\rho_\sigma$.
Thus, a  residual  isocurvature perturbation $\S_i$ will
inevitably be present at some level unless the relevant number
density $n_i$ is created after the curvaton has decayed.

In the sudden-decay approximation, it is easy to arrive at a
general formula for the residual isocurvature  perturbations. Note
first that before curvaton decay, $\rho_r$ is supposed to be
uniform on the flat slicing. The perturbation of the number
density $n_i(\rho\sub r,\rho_\sigma)$ on this slicing is therefore
just given by
 \be
 \delta n_i = \(\frac{\partial n_i}{\partial \rho_\sigma}
\) \delta \rho_\sigma \,.
 \ee
As a result the perturbation, \eq{tzi}, can be written as
 \be
 \tilde \zeta_i =\frac13 \(\frac{\partial n_i}{\partial \rho_\sigma}\)
  \frac{\delta \rho_\sigma}{n_i} \,.
 \ee
Using \eq{defr} for the total curvature perturbation after
curvaton decay, we have
 \be
 \frac{\tilde \zeta_i}{\zeta} =
\frac1{r} \frac{\rho_\sigma}{n} \(\frac{\partial n_i}{\partial
\rho_\sigma} \) \label{zetaigeneral} \,.
 \ee
This expression can be evaluated at any epoch
after the conservation of $n_i$ but before
curvaton decay, since both $n_i$ and $\rho_\sigma$ are
proportional to $a^{-3}$.
Courtesy of the separate universe assumption, its evaluation
requires only a knowledge of the function $n_i(\rho\sub
r,\rho_\sigma)$ in an unperturbed universe. This is similar in
spirit to the calculation in \cite{ewanmisao} of the curvature
perturbation produced by multi-field inflation. The separate
universe assumption can be verified by explicit calculation in all
three of the CDM cases that we shall deal with in the next
section, since the situation there is sufficiently simple, just as
it can be verified in the case of multi-field inflation.

In the case where the CDM or baryon number is created before
the curvaton decays, the residual CDM or baryonic matter
isocurvature perturbation
is  thus given by \eq{residual} with
 \bea
s\cdm
 &=& 3 \[ \frac 1r  \frac{\rho_\sigma}{n\cdm}
\(\frac{\partial n\cdm}{\partial \rho_\sigma} \) - 1 \]
 \label{cdmgeneral}\\
s_B &=& 3 \[
 \frac 1r
 \frac{\rho_\sigma}{n_B} \(\frac{\partial n_B}{\partial \rho_\sigma}\)
-1 \]
\,.
 \eea
In the opposite case where the curvaton decays and its decay products are
thermalised before the CDM or baryon number is created,
the CDM or
baryon number at creation is simply a function of the local
energy density, as in the inflaton scenario, and we have $s_i=0$.

\section{The residual CDM isocurvature perturbation}

\label{residualsec}

For the rest of this paper, we focus on the residual CDM
isocurvature perturbation, which is present in the curvaton
scenario if the CDM is created before the curvaton has decayed. We
shall work out \eq{cdmgeneral} in the cases  that the CDM  (i) is  a
WIMP (ii) is  an axion and (iii) comes from an oscillating field with
constant mass, such as a Wimpzilla. We shall also compare our
results with present observational bounds.

Before looking at the individual cases though, we
recall two results already considered in
\cite{luw}. The first result holds in the case that the CDM is
created at an epoch when $(\rho_\sigma/\rho)$ is negligible
compared with the value when the curvaton decays, which means in
particular that the CDM must be created well before the curvaton
decays. If the curvaton density has a negligible effect on
$n\sub{cdm}$, \eq{cdmgeneral} becomes
 \be
s\cdm  = - 3
\label{Scdmzeta}
 \,.
 \ee
This is too big to be compatible with the current observational bound
given in \eq{scdmbound}, and is independent of the nature of the CDM.
It follows that in the curvaton scenario, CDM cannot be created
significantly before the curvature perturbation achieves its final
value. In particular, CDM cannot be created just after inflation, as
would be the case if the CDM consisted of heavy weakly interacting
particles created from the vacuum. (This kind of CDM was originally
envisaged in \cite{dv,lrs}, and later \cite{wimpzillas} called
Wimpzilla CDM.)

The other result holds when the CDM is created by the decay of the
curvaton itself. The epoch of CDM creation then corresponds to the
epoch when the curvaton decay is complete. The resulting local CDM
density is a fixed multiple of the curvaton number density
well before decay. The fractional perturbations are thus equal and
hence
\be
\tilde\zeta\sub{cdm} = \zeta_\sigma \,.
\label{fromdecay}
\ee
In terms of $r$, defined in \eq{defr}, the prediction
\eq{fromdecay} leads to
\be
s\cdm
 =
3\( \frac{1-r}r \)
\,.
\ee
This is compatible with \eq{scdmbound} only if $1-r < 0.03 $, which
means that the curvaton has to dominate before it decays.  Note that
$r\leq1$ \cite{mwu} and hence in this case there is a positive
correlation between the primordial curvature perturbation $\zeta$ and
isocurvature perturbation $\S$, i.e., $s>0$.

\subsection{Weakly interacting massive particles}

If the CDM consists of weakly interacting massive particles (WIMPs),
it is initially in thermal equilibrium.  The number density of the
WIMPs is conserved only after the epoch when the WIMPs fall out of
thermal equilibrium (the epoch of freezout). In the present context
that epoch should be taken as the one when the CDM is created. In
contrast with the axion case that we shall discuss next, no scalar
field can be involved in the creation of WIMP CDM and therefore there
can be no intrinsic isocurvature perturbation.

The calculation of the residual isocurvature perturbation after
freeze-out in terms of gauge-invariant linear perturbations
is given in an Appendix~\ref{append}. However we shall show here how
the same result can be obtained working only in terms of the
unperturbed background equations, via \eq{cdmgeneral}.

In either case we work in the approximation of sudden freeze-out,
denoting the moment of freeze-out by a star. This moment is determined
by an equation of the form
\be
\label{defgamma}
\Gamma(T_*)/H_* = K
\,,
\ee
where $K$ is of order 1 and $\Gamma$ is the
interaction rate per CDM particle. The temperature dependence of the
interaction rate given in terms of a dimensionless parameter
\be
\alpha \equiv \frac{d\ln\Gamma}{d\ln T}
 \label{talpha}
\,,
\ee
with $\alpha>2$ if there is initially thermal equilibrium at high
temperatures.

The number density of non-relativistic particles ($m>T_*$) at
freeze-out is
\be
n_* = g_* \left( \frac{mT_*}{2\pi} \right)_*^\frac32 e^{-m/T_*}
\label{nfreeze}
\,,
\ee
where $m$ is the mass and $g_*$ the internal degrees of freedom of the
WIMP. 
The annihilation cross-section for non-relativistic WIMPs is weakly
dependent on temperature, with $\langle \sigma_A v \rangle\propto
T^{p/2}$ where $p=0$ for $s$-wave annihilation and $p=2$ for $p$-wave
annihilation \cite{KT}. The rate of change of the interaction rate
$\Gamma$ is thus mainly determined by the rapidly decreasing number
density if $m>T$, yielding
\be
\Gamma = n\langle \sigma_A v \rangle \propto T^{3/2+(p/2)} e^{-m/T}
 \,,
\ee
and hence $\alpha=(m/T)+(3/2)+(p/2)$.

After freeze-out interactions are negligible and the number density is
diluted as the universe expands, so we have $n=n_*(a_*/a)^3$.
During any era when the number of effective species is constant,
we have $a\propto 1/T$ and hence
\be
\label{defnT}
n= n_* (T/T_*)^3 = g_* (m/2\pi T_*)^\frac32 e^{-m/T_*} T^3
\,,
\ee

This is the expression that we want to insert into
\eq{cdmgeneral}. The partial derivative is at fixed $\rho_r$ and
therefore at fixed temperature, $T<T_*$, so that \eq{cdmgeneral}
can be written
\be
 \label{residualscdm}
s\cdm + 3 = \frac3r \left( \frac{\rho_\sigma}{n}
\frac{\partial n }{\partial \rho_\sigma} \right)_T \,. \ee
The local CDM number density after decoupling varies with the local
curvaton density because the curvaton density at freeze-out,
$\rho_{\sigma*}$, affects the expansion rate, $H_*$ in
\eq{defgamma}, and hence the local freeze-out temperature,
$T_*$. Hence we can write
\be
\label{step1}
\left( \frac{\rho_\sigma}{n}
\frac{\partial n }{\partial \rho_\sigma} \right)_T
 =
\(\frac{m}{T_*}-\frac32 \)  \left( \frac{\rho_\sigma}{T_*}
\frac{\partial T_*}{\partial \rho_\sigma} \right)_T
\ee
where we have used \eq{defnT} to obtain
$(\partial n/\partial T_*)_T=(m/T_*^2-3/2T_*) n$.

Assuming that the curvaton is oscillating at freeze-out (i.e.,
the curvaton mass is greater than the Hubble rate, $m_\sigma>H_*$) we
have
\be
\rho_\sigma = \rho_{\sigma*} \left( \frac{a_*}{a} \right)^3 =
\rho_{\sigma*} \left( \frac{T}{T_*} \right)^3 \,,
\ee
and hence
\be
\label{step2}
\left( \frac{\rho_\sigma}{T_*}
\frac{\partial T_*}{\partial \rho_\sigma} \right)_T
 =
 \left( \frac{T_*}{\rho_\sigma} \frac{d\rho_{\sigma*}}{dT_*} -3
 \right)^{-1}
\,.
\ee
Finally then we must determine $d\rho_{\sigma*}/dT_*$ from
\eq{defgamma}, which gives
%
\bea
dK
 &=& \left[ \alpha_* \frac{dT_*}{T_*} - \frac12\frac{d\rho_*}{\rho_*}
 \right] \frac{\Gamma(T_*)}{H_*} \,,\nonumber\\
&=&
\left[ \left( \alpha_*-2(1-\Omega_{\sigma*}) \right) \frac{dT_*}{T_*}
 - \frac12 \Omega_{\sigma*}
 \frac{d\rho_{\sigma*}}{\rho_{\sigma*}}
 \right]
\frac{\Gamma(T_*)}{H_*} \,,\nonumber\\
&=& 0 \,.
\eea
and hence
%
\be
\label{step3}
 \frac{T_*}{\rho_\sigma} \frac{d\rho_{\sigma*}}{dT_*} =
  \frac{2(\alpha_*-2)+4\Omega_{\sigma*}}{\Omega_{\sigma*}}
\,.
\ee

Combining \eqsss{residualscdm}{step1}{step2}{step3} we then obtain the
residual isocurvature perturbation,
\be
s\cdm =
3 \[  \frac {\Omega_{\sigma *} }{ r }
\( \frac{m}{T_*}-\frac32 \)
\frac1 {2(\alpha_*-2) + \Omega_{\sigma *} } - 1
\]
\,.
\label{finalwimp}
\ee
%

In the usual case that freeze-out occurs during radiation domination
(requiring $\Omega_{\sigma*}\ll1$) with the WIMP a neutralino, then
$m/T_*$ is of order 20.
In this case the annihilation cross-section is
approximately constant while the number density drops rapidly and
hence
\be
\alpha_* \approx \( \frac{d\ln n}{d\ln T} \)_* \approx \frac{m}{T_*} \,,
\ee
and \eq{finalwimp} reduces to
\be
s\cdm \approx
 3 \[  \frac {\Omega_{\sigma *}}{2r} - 1 \]
\,.
\label{finalwimp2}
\ee
The first term in the bracket is thus roughly of order $\Omega_{\sigma
  *}/\Omega_{\sigma {\rm\,dec}}$ (using the sudden-decay approximation
\eq{suddec} which gives $r\sim\Omega_{\sigma{\rm\,dec}}$).
If freezout occurs well before the curvaton decays and while the
curvaton accounts for only a small fraction of the energy density
(i.e., for $\Omega_{\sigma *}\ll\Omega_{\sigma {\rm\,dec}}$),
this makes $s\cdm \simeq -3$ in accordance with the general
expectation \cite{luw} for this regime, which violates the
observational bound.

On the other hand, if the curvaton dominates the
energy density before freeze-out so that $\Omega_{\sigma
  *}\sim\Omega_{\sigma {\rm\,dec}}\sim1$ (or if freeze-out occurs close
to curvaton decay with $\Omega_{\sigma *}\sim\Omega_{\sigma
  {\rm\,dec}}<1$), there could be a regime where $s\cdm$ is small
enough to satisfy observational bounds.
Note that if the CDM freezes-out when the curvaton has come to dominate
the energy density ($\Omega_{\sigma *}\simeq1$) but before it decays,
then the CDM abundance relative to the radiation density will be
significantly diluted when the curvaton decays. Thus $m/T_*$ could be
significantly smaller than usually assumed. In order to determine the
allowed regime more precisely would require more detailed numerical
modeling going beyond the sudden freeze-out and sudden decay
approximations that we have used here.

\subsection{Axion CDM}

If the CDM consists of axions, it is created when the rising axion mass
$m(T)$ becomes equal to the falling Hubble parameter $H$ because only
then does  the approximately homogeneous axion field become free to oscillate.
If axionic strings are present, the axions radiated from them  will
probably be the dominant creation mechanism. In that case there can be
no intrinsic CDM isocurvature perturbation. If instead such strings are
absent, the axion field $\chi(\bfx,t)$ oscillates
independently at each comoving
 position, with an initial value that is the same as it was during inflation.
(This is the separate universe assumption, which in the present case
can be verified by
explicit calculation \cite{myaxion} from the axion field equation.)
Then, the axion density is proportional to the square of the
initial axion field value,  and a perturbation in this value will generate
an intrinsic isocurvature perturbation
\be
 \S\cdm = \frac{2\delta\chi}\chi
\,.
\label{saxion}
\ee

We are here concerned instead with the
residual isocurvature perturbation, which is always present if the
curvaton decays after the epoch of axion creation.
We work in the approximation that the oscillation starts suddenly and is
immediately harmonic, which is known to be reasonable in most of parameter
space \cite{myaxion}.
Then, if axion creation occurs during radiation domination  the
calculation is very similar to the one that we gave for the WIMP case.
The freezout epoch is determined by
\be
m(T_*)/ H_* = K
\,,
\ee
with $K$ conventionally set equal to 1 exactly. The increasing mass is
given accurately by
\be
m\propto T_*^{-\beta}
\,,
\ee
with $\beta = -3.7 $ \cite{turner86}.
Assuming that there is no intrinsic perturbation,
the axion number density $n_*$ at creation is proportional to $m(T_*)$.
Repeating the calculation of the WIMP case we find for the axion case
\be
s\cdm = 3 \[
\frac{\Omega_{\sigma *}}{\Omega_{\sigma {\rm \, dec}} }
\frac{3+\beta}{2\beta +4 - \Omega_{\sigma *} } - 1 \]
\label{finalaxion}
\,.
\ee
In contrast with \eq{finalwimp}, this always gives $s\cdm \simeq
-3$.
 We conclude that if the CDM consists of axions, the
 curvaton cannot decay after  the CDM is created at $T\sim \GeV$.

\subsection{CDM from an  oscillating field of fixed mass}

The previous calculation does not apply if axion creation occurs
at low temperatures, and in particular if it occurs while the
curvaton density dominates. In such a case, the temperature of the
sub-dominant radiation will have a negligible effect on the axion
mass, which will therefore have its vacuum value.

More generally, the CDM might consist of particles corresponding
to the oscillation of some  scalar field other than the axion, in
which  case the mass will generally be at the vacuum value whether
or not creation takes place during matter domination. This
possibility is envisaged in the rather attractive proposal of
Moroi and Randall \cite{mr}, whereby a modulus decays before nucleosynthesis
into both baryonic matter and CDM. The decay occurs because the
modulus mass is $10$ to $100\TeV$ instead of the usual $1\TeV$ or
so. In contrast with earlier proposals though, the high mass is
explained naturally, arising from the fact that in this scenario
SUSY-breaking   is anomaly-mediated as opposed to
gravity-mediated. The anomaly-mediation also ensures that matter
will not be over-produced by the decay. Of course, this idea
has been worked out only under within the inflaton scenario and it
remains to be seen whether it can be viable in  the curvaton
scenario, with the oscillation starting before the curvaton
decays as we are envisaging.

Without focusing on any particular scenario, we consider the residual
CDM isocurvature perturbation that is produced if the CDM does originate from
the oscillation of some field with fixed mass.
With constant mass, corresponding to $\beta=0$, \eq{finalaxion} can be written
 \be
s\cdm =
3 \( \frac {f_*} { \Omega_{\sigma {\rm \,, dec}} } -1 \)
\label{fixedmass}
\,,
 \ee
where $f$ is defined in \eq{fdef}.
Using \eq{zetat} this  equation corresponds actually to
 \be
\tilde \zeta\cdm = \zeta_*
\,.
 \ee
This in turn is simply a consequence of the fact that the oscillation
starts on a slice of uniform energy density, corresponding to local
Hubble parameter $H=m$.

If  the oscillation starts well before the curvaton decays,  and
while  the curvaton accounts for only a small fraction of the
energy density, this again gives $s\cdm \simeq - 3$ in
contradiction with observation. However, in contrast with the
cases of  WIMP and axionic CDM, $s\cdm$ in the present case
vanishes in the limit that the curvaton completely dominates the
energy density when the CDM is created (ie.\ when the CDM
oscillation starts). Using \eq{fixedmass} one finds in this regime
 \be
s\cdm \simeq -\frac43 \( \frac {\rho\sub r }{\rho_\sigma} \)_* +
 \( \frac {\rho\sub r }{\rho_\sigma} \)\sub{dec}
\label{fixedmass2}
\,.
 \ee
Unfortunately, 
the sudden-oscillation and sudden-decay
approximations both become inadequate in this regime, so that
\eq{fixedmass2} will not in fact be correct. Departures in the
sudden-oscillation can be calculated once the potential of the
oscillating is known but no relevant calculation has been
published so far. Even in the absence of such a calculation
though, it follows just from continuity that that there will be a
regime in which $s\cdm$ is within the present observational bound,
while being big enough to be observable in the future.

\section{Conclusion}

\label{conclusion}

The observed primordial curvature perturbation may well be accompanied
by isocurvature perturbations in CDM, baryonic matter or neutrino. In
the curvaton scenario, such a perturbation is inevitably generated by
the curvaton perturbation, unless the curvaton decays before the
relevant quantity (CDM, baryon number or lepton number) is created.
This `residual' isocurvature perturbation is fully correlated with the
curvature perturbation.

We presented a general expression, \eq{zetaigeneral}, which allows the
residual isocurvature perturbation to be evaluated once the mechanism
for generating the relevant quantity is specified.  We then evaluated
the residual CDM isocurvature perturbation for all the commonly
considered candidates in the sudden decay approximation, and compared
it with the observational upper bound
\cite{gl}.  We first recalled previously-known results \cite{luw}. The
CDM cannot be created right after inflation in the curvaton scenario,
as in the case of CDM in the form of super-heavy weakly interacting
massive particles (Wimpzillas), created from the vacuum energy
\cite{lrs,wimpzillas}. But if the curvaton dominates the energy
density before it decays, then the CDM can be created from that decay.

We then went on to consider new cases, using \eq{cdmgeneral}.
After a fairly complicated calculation, we derive \eq{finalwimp}
for the case of CDM falling out of thermal equilibrium (weakly
interacting massive particle or WIMP CDM). It is compatible with
observation only if the CDM is created when the curvaton density
has a rather specific value. Modifying the previous  calculation
slightly, we arrive at the case of axion CDM given by
\eq{finalaxion}, which we find is too big in all cases. These
results mean that axions, and probably also WIMPS,  have to
be created {\em after} curvaton decay. Such a requirement is not
however a strong constraint on the curvaton scenario, since
WIMPS and axions are both typically created rather late (at
temperatures respectively of order $10\GeV$ and $1\GeV$).

Finally, we consider the case of CDM created from the oscillation
of a scalar field with fixed mass. As the created particle would
need to be massive and very weakly interacting to avoid subsequent
thermalisation, we are again dealing with a Wimpzilla. We derive
\eq{fixedmass} for the residual isocurvature perturbation in this
case. It again gives $s\cdm =-3$ in the limit where the curvaton
density  when the oscillation starts is much smaller than its
final value. In contrast with the previous cases though, it gives
$s\cdm=0$ in the opposite limit where curvaton domination is
complete when the oscillation starts. There must therefore be an
intermediate regime where $s\cdm$ in this case is of potentially
observable magnitude.

We can summarise the situation as follows. The residual CDM
isocurvature perturbation is absent if the CDM is created after
curvaton decay. Otherwise it is present, and it is typically either
too big to be compatible with observation, or completely negligible.
There are cases though where it can be of observable magnitude. In
particular, this can occur if the CDM comes from the decay of the
curvaton, or from the oscillation of a scalar field of fixed mass.  In
both of these cases, the residual isocurvature perturbation vanishes
in the limit that the curvaton completely dominates the energy density
at the time of the CDM creation, but isocurvature perturbations can
exist at a detectable level if the domination is not complete.

\acknowledgments We are grateful to Y.\ C.\ Kim for a comment
on our treatment of the WIMP case.
This work was supported in
part by PPARC grants PPA/G/S/1999/00138, 
PPA/G/S/2000/00115 and 
PGA/G/O/2000/00466.  DW is
supported by the Royal Society.


\appendix

\section{Gauge-invariant calculation}

\label{append}


In this section we reproduce the calculation of the CDM perturbation
given in section \ref{residualsec} in terms of manifestly gauge
invariant perturbations.

To determine the CDM isocurvature perturbation we need to determine
the gauge-invariant perturbation in the the CDM number density on
spatially flat slices, \eq{tzi}, (or equivalently the curvature
perturbation on uniform number-density slices)
\be
  \label{defzetaDW}
 \tilde\zeta\cdm
 \equiv {1\over3}{\delta n\cdm\over n\cdm} - \psi\,.
 \ee
where $\delta n\cdm$ and $\psi$ are the gauge-dependent
perturbed number density and curvature perturbation respectively.

In the {\em sudden freeze-out} approximation, we assume that the
CDM remains in equilibrium abundance, with number density given by
\eq{nfreeze},
until the local interaction rate per expansion time falls below the
critical value
\be \label{fo}
 \left. {\Gamma \over H} \right|_f = K \,.
 \ee
Thereafter the CDM number is conserved and hence the curvature
perturbation $\tilde\zeta\cdm$ in \eq{defzetaDW} remains constant on
large scales after freeze-out.
The physical condition (\ref{fo}) picks out a physical
hypersurface $\Sigma_f$ which marks the transition between
equilibrium and freeze-out. Thus we can determine $\tilde\zeta\cdm$ in
Eq.~(\ref{defzetaDW}) after freeze-out from the curvature
perturbation, $\psi_f$, and perturbed number density, $\delta
n_f$, on this hypersurface.

A uniform temperature slice at this time $\Sigma_T$ has no
perturbation in the equilibrium number density, $\delta n_T=0$,
but may have a curvature perturbation $\psi_T=-\zeta_r$. (In the
curvaton scenario we have $\zeta_r\ll\zeta_\sigma$ before the
curvaton decays.)
We obtain $\psi_f$ and $\delta n_f$ via a gauge-shift from the
hypersurface $\Sigma_T$ to the hypersurface $\Sigma_f$. This
corresponds to a coordinate-shift
\be
 \label{firsttfr}
 \delta t_{fT} = {\delta (\Gamma/H)_T \over
 (\dot\Gamma/H)-(\dot{H}\Gamma/H^2)} \,.
\ee
We then have
\bea
 \psi_f &=& \psi_T + H\delta t_{fT} \,, \\
 \delta n_f &=& \delta n_T - \dot{n}\delta t_{fT} \,,
\eea
and hence
\be
 \label{tz2}
 \tilde\zeta\cdm = \zeta_r + \frac13 \left( \frac{m}{T_*} -\frac32
 \right) H_* \delta t_{fT} \,,
\ee
where we have assumed $H^{-1}\dot{T}/T=-1$.

Assuming $\Gamma/H\propto T^\alpha/\rho^{1/2}$ (where $\alpha>2$ for
CDM to fall out of equilibrium with decreasing temperature)
we have from \eq{firsttfr}
\be
 \delta t_{fT} = \left(
 \frac{4-\Omega_\sigma}{2(2-\alpha)-\Omega_\sigma} \right)_*
 \frac{\delta\rho_T}{\dot\rho_*} \,,
\ee
and using
\be
 {H\delta\rho_T \over \dot\rho} = \zeta_r-\zeta = {3\Omega_{\sigma}
 \over 4-\Omega_{\sigma}}
  (\zeta_r-\zeta_\sigma)\,,
\ee
we then have
\be
 \label{secondtfr}
 H_* \delta t_{fT} = \left(
 \frac{3\Omega_{\sigma*}}{2(\alpha-2)+\Omega_{\sigma*}} \right)_*
 (\zeta_\sigma-\zeta_r) \,.
\ee

Finally we then obtain from \eq{tz2} and (\ref{secondtfr})
\be
 \label{tzresult}
 \tilde\zeta\cdm
 = \zeta_r + \left(\frac{m}{T_*}-\frac32\right) {\Omega_{\sigma*}
 \over 2(\alpha-2)+\Omega_{\sigma*}} (\zeta_\sigma-\zeta_r) \,.
\ee

If we have adiabatic perturbations ($\zeta_r=\zeta_\sigma$) before
freeze-out, we must have $\tilde\zeta\cdm=\zeta_r$ from \eq{tzresult},
and hence $s\cdm=0$ after freeze-out.

In the curvaton scenario when the CDM decouples before the curvaton
decays, we have $\zeta_\sigma\gg\zeta_r$ and the primordial curvature
perturbation is given by $\zeta=r\zeta_\sigma$.  The residual CDM
isocurvature perturbation is then given by \eq{finalwimp} of
section~\ref{intres}.



\end{document}